\documentclass[
    ,final            
  ]
  {aipproc}

\layoutstyle{6x9}
\usepackage{amsmath}
\usepackage{array}
\usepackage{float}	
\usepackage{amssymb}
\usepackage{color}

\begin{document}

\title{Impact of r-modes on the cooling of neutron stars}
 

\author{Mark G. Alford}{
  address={Department of Physics, Washington University in St. Louis, Missouri, 63130, USA}
}

\author{Simin Mahmoodifar}{
  address={Department of Physics, University of Maryland
College Park, MD 20742, USA}
}

\author{Kai Schwenzer}{
  address={Department of Physics, Washington University in St. Louis, Missouri, 63130, USA}
}

\begin{abstract}
Studying the frequency and temperature evolution of a compact star can give us valuable information about the microscopic properties of the matter inside the star. In this paper we study the effect of dissipative reheating of a neutron star due to r-mode oscillations on its temperature evolution. We find that there is still an impact of an r-mode phase on the temperature long after the star has left the instability region and the r-mode is damped completely. 
With accurate temperature measurements it may be possible to detect this trace of a previous r-mode phase in observed pulsars.  \end{abstract}

\maketitle



\section{Introduction}

Neutron stars are the remnants of the core collapse supernova explosions of massive stars which are born with temperatures higher than $10^{11}\,$K. 
For core temperatures larger than $10^6$-$10^7\,$K the cooling of the star is dominated by neutrino emission from its interior. Oscillations of the star, such as r-mode oscillations, also affect the temperature evolution of the star by viscous heating which can drastically alter the thermal evolution.

The r-modes are non-radial pulsations of neutron stars that are primarily driven by Coriolis forces and are coupled to gravitational radiation \cite{Friedman:1978hf,Andersson:1997xt,Friedman:1997uh}. R-modes are 
damped by bulk and shear viscosities \cite{Lindblom:1998wf}, and therefore they connect the microscopic properties of the matter inside the star, which depend on the low energy degrees of freedom and the equation of state, to the macroscopic and observable properties of the star. 
R-modes are
generally defined to be the solutions of the perturbed fluid equations having (Eulerian) velocity perturbations of the form 
\\
\begin{equation}
\delta \vec{v}=\alpha R \Omega \left(\frac{r}{R} \right)^m \vec{r} \times \vec{\nabla}Y_{mm} e^{i\omega t}+O(\Omega^3)
\end{equation}
where $\alpha$ is a dimensionless amplitude, $R$ and $\Omega$ are the radius and angular velocity of the unperturbed star, $Y_{mm}$ is the spherical harmonic and $\omega$ is the frequency of the mode, given by $\omega=-(m-1)(m+2)/(m+1)\Omega + O(\Omega^3)$ in the inertial frame.
The r-modes evolve with time dependence of $e^{(i\omega t -\frac{t}{\tau})}$, where $1/\tau$ can be decomposed as

\begin{equation}
\frac{1}{\tau(\Omega)}=\frac{-1}{|\tau_{G}(\Omega)|}+\frac{1}{|\tau_{B}(\Omega)|}+\frac{1}{|\tau_{S}(\Omega)|} +\cdots
\end{equation}
Here $\tau_{G}, \tau_{B}$ and $\tau_{S}$ are gravitational radiation, bulk viscosity and shear viscosity timescales, respectively, and the dots denote other dissipative mechanisms.
R-modes are unstable when $\tau<0$.

A phenomenological model describing the r-mode evolution has been given by Owen
et. al. \cite{Owen:1998xg} (see also \cite{Ho:1999fh}). Since the amplitude of the
r-mode grows exponentially in the unstable region, where the gravitational
radiation timescale is smaller than the damping timescale due to viscosity, there
must be some non-linear mechanisms that saturates the amplitude at a finite
value. Supra-thermal bulk viscosity \cite{Alford:2010gw} is one of these non-linear
mechanisms, but in the case that only the core of the star is considered and the
effect of the crust is ignored, it can only saturate the r-mode at large
amplitudes \cite{Alford:2011pi}. Mode coupling
is another mechanism that can saturate the r-mode at small amplitude when
hyperons and therefore large hyperon bulk viscosity is included in the
computations \cite{Arras:2002dw, Bondarescu:2008qx}. Core-crust boundary layer effects \cite{Wu:2000qy,Bildsten:2000ApJ...529L..33B} are another possible
mechanism. At this point it is not entirely clear which mechanism is actually responsible for
saturating the r-mode amplitude.  In this paper we study the evolution of a neutron star in the presence of r-mode oscillations with focus on the effects on its thermal evolution. Our approach in studying the evolution of
the star is similar to \cite{Yang:2009iu}, but the main difference that distinguishes our
results from previous analyses is a difference in the reheating term which
appears in the temperature evolution.


%
%

\section{R-mode evolution}

Studying the spin-down and temperature evolution of a neutron star that involves r-modes requires to solve a system of three coupled evolution equations for the rotation frequency $\Omega$ of the star and the r-mode amplitude $\alpha$, as well as the temperature $T$. These equations are derived from energy and angular momentum conservation \cite{Ho:1999fh}
\begin{equation}
\frac{d\Omega}{dt}=-2Q\frac{\Omega\alpha^2}{\tau_V} \; ,\quad
\frac{d\alpha}{dt}=-\frac{\alpha}{\tau_G}-\frac{\alpha}{\tau_{V}}(1-\alpha^2Q) \; , \label{eq:evolution}
\end{equation}
where $Q\equiv 3\tilde{J}/(2\tilde{I})$
and the damping time $\tau_V$ is given by $1/\tau_V=1/\tau_S + 1/\tau_B + \cdots$ . The dots denote other possible dissipative mechanisms, like boundary layer effects. The time scales $1/\tau_i$ are given by
\begin{equation}
\frac{1}{\tau_i}=-\frac{P_i}{2E_m} \; ,
\end{equation}
where $P_G$ is the power radiated by gravitational waves and $P_B$ and $P_S$ are dissipated powers due to (subthermal) bulk and shear viscosity, and $E_m$ is the energy of the mode

\begin{align}
P_G &=\frac{32\pi(m-1)^{2m}(m+2)^(2m+2)}{((2m+1)!!)^2(m+1)^{2m+2}}\tilde{J}_m^2GM^2R^{2m+2}\alpha^2\Omega^{2m+4} \; , \\
P_B &=-\frac{16m}{(2m+3)(m+1)^5\kappa^2}\frac{\tilde{V}_m\Lambda_{{\rm QCD}}^{9-\delta}R^7\alpha^2\Omega^4T^\delta}{\Lambda_{EW}^4} \; , \label{eq:bulk-dissipation} \\
P_S &=-\frac{(m-1)(2m+1)\tilde{S}_m\Lambda_{{\rm QCD}}^{3+\sigma}R^3\alpha^2\Omega^2}{T^\sigma} \; , \\
E_m &=\frac{1}{2}\alpha^2\Omega^2MR^2\tilde{J}_m \; .
\end{align}
Here we concentrate on the lowest $m=2$ r-mode whereas $\delta$ and $\sigma$ are given in Table~\ref{tab:viscosity-parameters}. The parameters $\tilde{V}$, $\tilde{S}$, $\tilde{I}$ and $\tilde{J}$, which are given in Tables~\ref{tab:viscosity-parameters} and ~\ref{tab:heating-cooling-parameters}, encode the radial integration over the star and $\Lambda_{{\rm QCD}}$ and  $\Lambda_{{\rm EW}}$ are characteristic strong and electroweak scales introduced to make these quantities dimensionless. In our calculations we have used $\Lambda_{QCD}=1$ GeV and $\Lambda_{EW}=100$ GeV.

\begin{table}
\begin{tabular}{|c|c|c|c|c|c|c|c|}
\cline{1-8} 
Neutron Star & Shell & $R[km]$&$\Omega_K[Hz]$& $\tilde{S}$ & $\tilde{V}$&$\sigma$&$\delta$\tabularnewline
\cline{1-8} 
NS $1.4\, M_{\odot}$ & core & $11.5$&$6020$& $7.68\times10^{-5}$ & $1.31\times10^{-3}$&$\frac{5}{3}$&$6$ \tabularnewline
\cline{1-1} \cline{3-6} 
NS $2.0\, M_{\odot}$ &  & $11.0$&$7670$& $2.25\times10^{-4}$ & $1.16\times10^{-3}$& & \tabularnewline
 \cline{1-6} 
NS $2.21\, M_{\odot}$ & m.U. core &$10.0$&$9310$& $5.05\times10^{-4}$ & $9.34\times10^{-4}$& & \tabularnewline
\cline{2-2} \cline{6-8} 
 & d.U. core & & & &$1.16\times10^{-8}$&$\frac{5}{3}$&$4$ \tabularnewline
\cline{1-8} 
\end{tabular}

\caption{\label{tab:viscosity-parameters}Radius, Kepler frequency and radial integral parameters that appear in the dissipative powers due to shear and bulk viscosity for different neutron star masses \cite{Alford:2010fd}.}

\end{table}
\begin{table}
\begin{tabular}{|c|c|c|c|c|c|c|c|}
\cline{1-8} 
Neutron Star & Shell & $\tilde{I}$& $\tilde{J}$ & $\tilde{C}_V$ & $\tilde{L}$&$\upsilon$&$\theta$\tabularnewline
\cline{1-8} 
NS $1.4\, M_{\odot}$ & core & $0.283$ & $1.81\times10^{-2}$ & $2.36\times10^{-2}$ & $1.91\times10^{-2}$&$1$&$8$ \tabularnewline
\cline{1-1} \cline{3-6} 
NS $2.0\, M_{\odot}$ &  &$0.300$&$2.05\times10^{-2}$ & $2.64\times10^{-2}$ & $1.69\times10^{-2}$& & \tabularnewline
 \cline{1-6} 
NS $2.21\, M_{\odot}$ & m.U. core &$0.295$& $2.02\times10^{-2}$ & $2.62\times10^{-2}$ & $1.29\times10^{-2}$& & \tabularnewline
\cline{2-2} \cline{6-8} 
 & d.U. core & & & &$2.31\times10^{-5}$&$1$&$6$ \tabularnewline
\cline{1-8} 
\end{tabular}

\caption{\label{tab:heating-cooling-parameters}Radial integral parameters
that appear in the moment of inertia, angular momentum of the mode, specific heat and neutrino emissivity of the star.}

\end{table}


The third equation for the temperature evolution is obtained by noting that the temperature of the star decreases mainly due to neutrino emission from the interior which in an average mass hadronic star is dominated by modified Urca process (in a massive star direct Urca process are allowed in an inner core and should be included in the neutrino emissivity as well), and it increases due to the dissipation of the r-mode energy by viscosity and other mechanisms


\begin{equation}
\frac{dT}{dt}=-\frac{1}{C_V} \left( L_{\nu}+P_V \right) \; , \label{eq:T-evolution}
\end{equation}
where the dissipated power is $P_V = P_S + P_B + \cdots$. Here the total (subthermal) neutrino luminosity $L_{\nu}$ and the integrated specific heat $C_V$ of the star are given by

\begin{equation}
L_{\nu} =\frac{4\pi R^3\Lambda_{QCD}^{9-\theta}\tilde{L}}{\Lambda_{EW}^4}T^{\theta} \; , \quad C_V =4\pi\Lambda_{QCD}^{3-v}R^3\tilde{C}_VT^\upsilon \; . \label{eq:neutrino-emissivity} \end{equation}
The dimensionless parameters $\tilde{L}_{\nu}$ and $\tilde{C}_V$ that involve radial integration over the star and $\theta$ and $v$ are given in Table ~\ref{tab:heating-cooling-parameters}.

%
%


Since we don't know which mechanism is actually responsible for the saturation of the r-mode amplitude we employ the simple model used in \cite{Owen:1998xg}. When solving the evolution equations we assume that there is a non-linear dissipative mechanism that saturates the mode at a fixed value $\alpha_{\rm sat}$. The evolution is then performed in three steps. First the evolution equations (\ref{eq:evolution}) and (\ref{eq:T-evolution}) are solved until the amplitude reaches its saturation value $\alpha_{\rm sat}$. Saturation requires then that there is a strong amplitude-dependent dissipative mechanism so that the 
the right hand side of the amplitude equation in eq. (\ref{eq:evolution}) vanishes at saturation, so that
\begin{align}
\frac{1}{\tau_V} &=\frac{1}{\tau_G}\frac{1}{1-\alpha_{\rm sat}^2 Q}
\end{align}
and the amplitude evolution stops. Therefore at saturation, in the residual evolution equations $1/\tau_V$ has to be replaced accordingly and the required dissipation to saturate the mode is entirely determined by gravitational physics. Since the dissipation that stops the growth of the mode inevitably heats the star, in contrast to \cite{Owen:1998xg} we also consistently replace $P_V$ by $P_G/(1-\alpha^2 Q)$ in the thermal equation.
When viscosity alone can not stop the growth of the r-mode and an additional dissipative saturation mechanism is required this leads to a considerably stronger reheating than when only the effect of viscous dissipation is considered, as has been done in previous analyses \cite{Yang:2009iu,Drago:2007iy,Andersson:2001ev}. The evolution in this saturation stage is performed until the boundary of the instability region is reached. From that point on, in a third step, the initial system of equations is solved to describe the subsequent decay of the r-mode and the cooling outside of the instability region.

Here we study the evolution of a neutron star that is made of ``APR hadron matter'' \cite{Akmal:1998cf} (see \cite{Alford:2010fd} for more details on our model). We also assume that the core of the star dominates the relevant quantities and neglect the effect of the crust.
For our star model shear viscosity arises from leptonic scattering and the bulk viscosity and neutrino emission is dominated by modified Urca reactions. 


\begin{figure*}
\begin{minipage}[t]{0.5\textwidth}%
\includegraphics[width=\hsize]{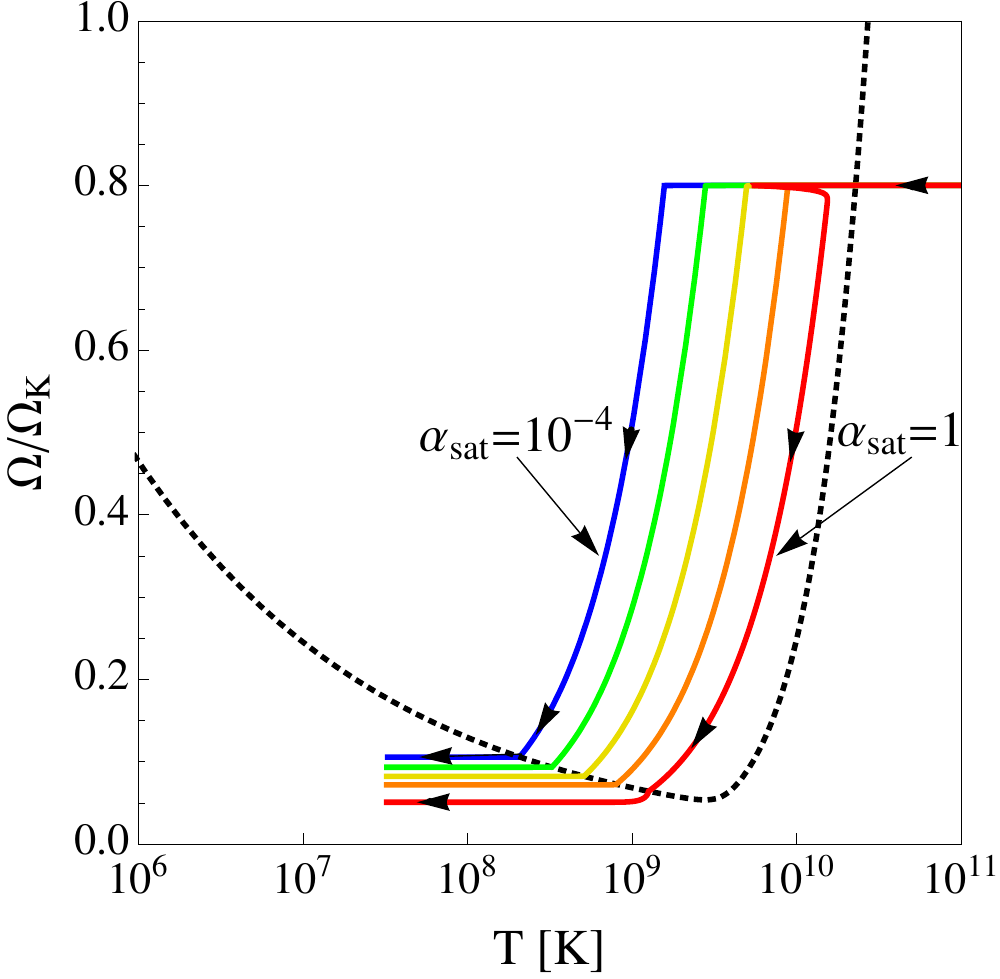}%
\end{minipage}%
\begin{minipage}[t]{0.5\textwidth}%
 \includegraphics[width=0.98\hsize]{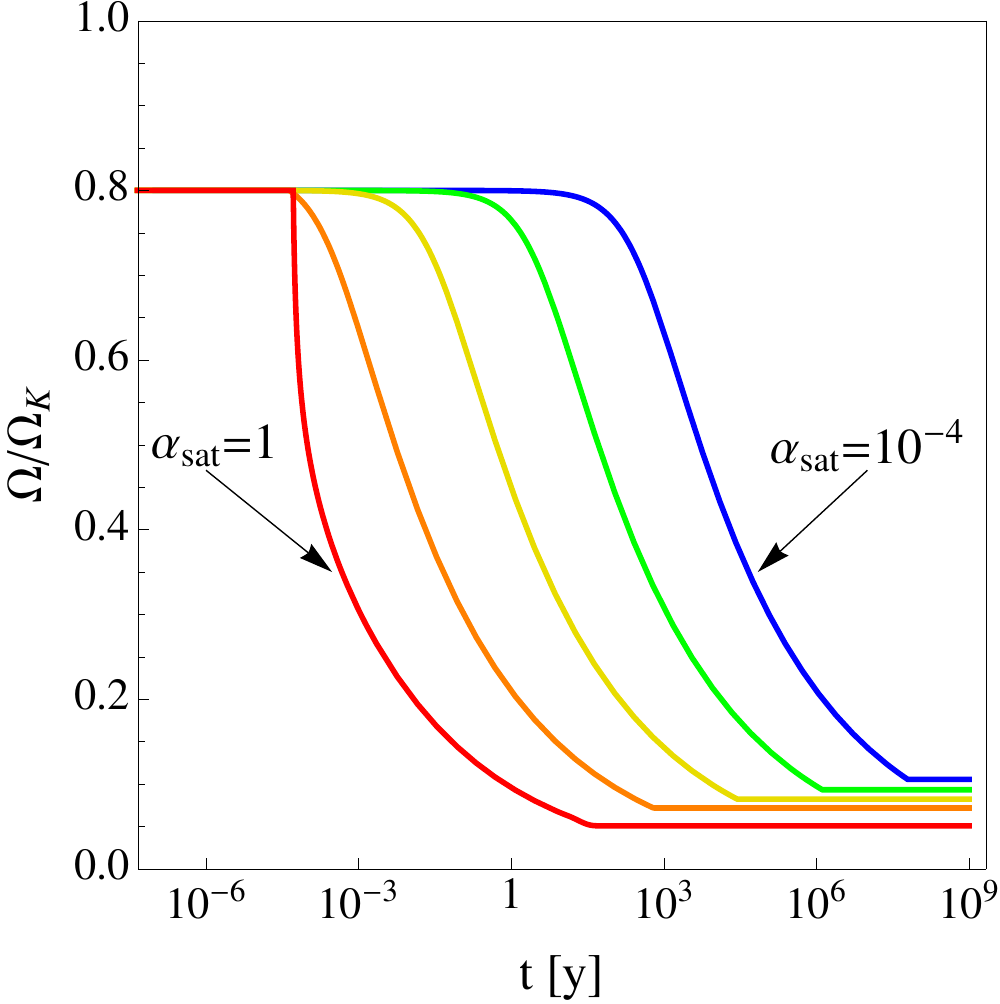}%
\end{minipage}
\caption{\label{fig:Omega} {\em Left panel:}
The spindown evolution of a young $1.4\,M_{\odot}$ APR neutron star in the angular frequency versus temperature plane. Different curves correspond to different saturation amplitudes, from $10^{-4}$ on the left to $1$ on the right. The dotted curve shows the boundary of the instability region. The initial conditions that we used for solving the evolution equations are $\Omega_0=0.8\,\Omega_K$, where $\Omega_K$ is the Kepler frequency, $\alpha_0=10^{-6}$ and $T_0=30\,{\rm MeV}$. {\em Right panel:} The evolution of the angular frequency of the star as a function of time for the same model shown on the left panel.}
\end{figure*}

\begin{figure*}
\begin{minipage}[t]{0.5\textwidth}%
\includegraphics[width=0.95\hsize]{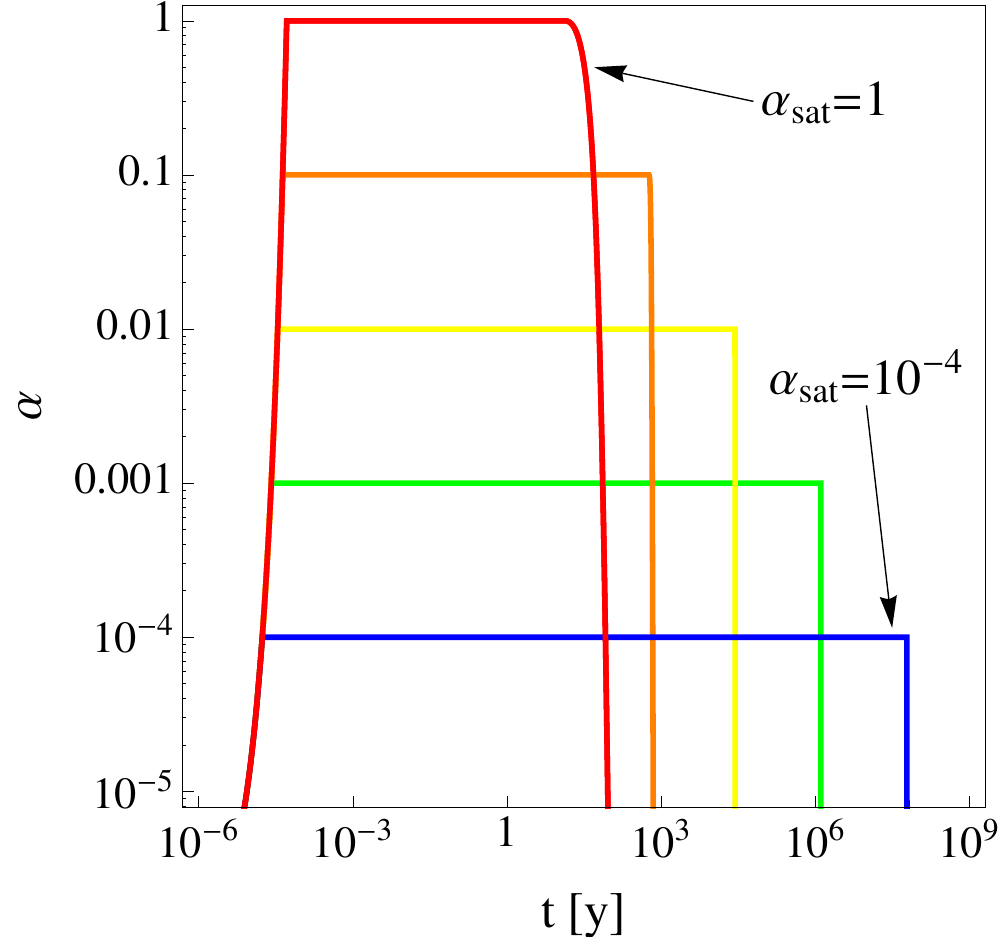}%
\end{minipage}%
\begin{minipage}[t]{0.5\textwidth}%
\includegraphics[width=\hsize]{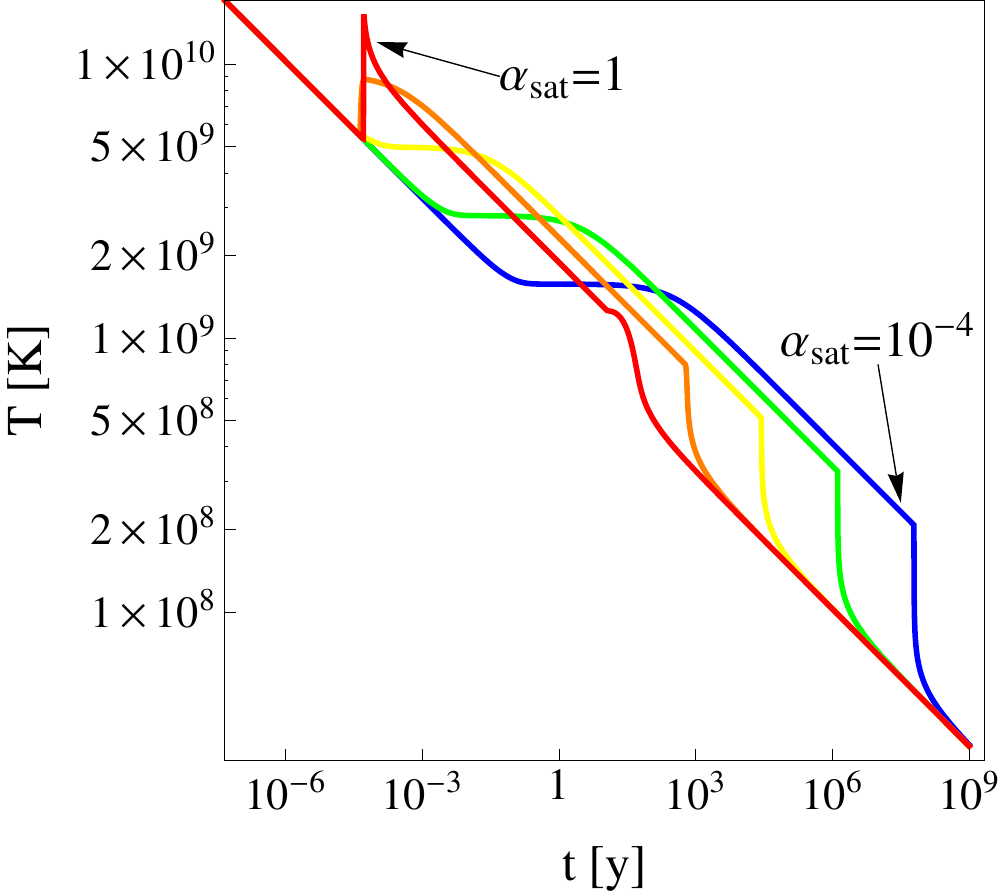}%
\end{minipage}

\caption{\label{fig:alpha-and-T}
{\em Left panel:}
Evolution of the r-mode amplitude with time for the same model as Fig.\ref{fig:Omega}; {\em right panel:} Evolution of the temperature of the star with time.
}
\end{figure*}

%
%

\section{Results and Discussion}

\begin{figure*}

\includegraphics[width=0.5\hsize]{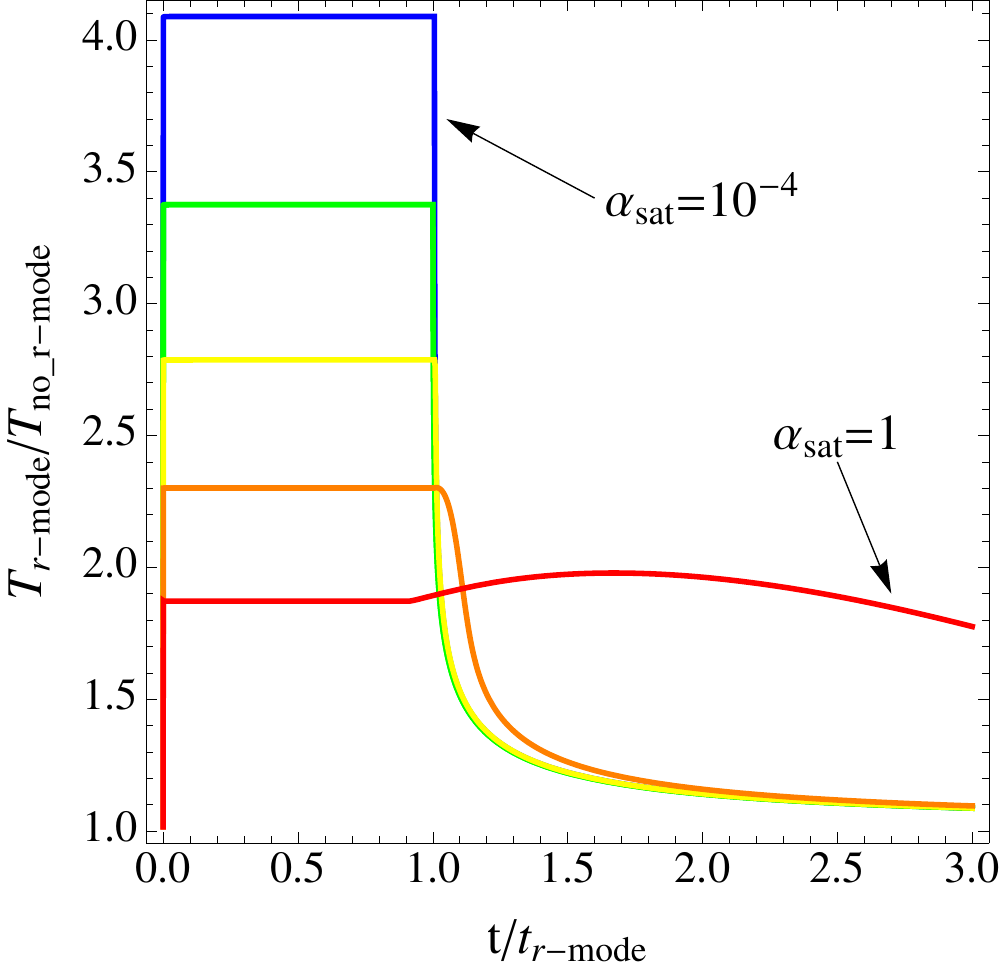}%

\caption{\label{fig:T-ratio}
Ratio of the star's temperature with and without the reheating due to r-mode dissipation versus time in units of the r-mode spin-down time (which increases strongly with decreasing saturation amplitude as seen in fig. \ref{fig:alpha-and-T}).
}
\end{figure*}

In Figs. \ref{fig:Omega}, \ref{fig:alpha-and-T} and \ref{fig:T-ratio} we show the results for the evolution of a young $1.4\,M_{\odot}$ APR neutron star with the following initial conditions: $\Omega_0=0.8\,\Omega_K$, where $\Omega_K=4/9\sqrt{2\pi G\rho_0}$ is the Kepler frequency, $\alpha_0=10^{-6}$ and $T_0=30\,\rm{MeV}$. We perform our calculations for different values of $\alpha_{\rm sat}$, from $10^{-4}$ to $1$. 
The left panel of Fig~\ref{fig:Omega} shows the rotation frequency of the star versus temperature. The dotted curve in that plot shows the boundary of the r-mode instability region for our star model where in the region above the curve the r-mode is unstable ($\tau_G < \tau_V$) and its amplitude grows exponentially until it gets saturated by the nonlinear mechanism, whereas in the region below the curve viscosity can damp the r-mode ($\tau_G > \tau_V$). As can be seen in Fig.~\ref{fig:Omega}, in the beginning of the evolution, cooling is very fast and the star enters the instability region in a fraction of a second. The amplitude of the r-mode grows exponentially until the amplitude reaches its saturation value $\alpha_{\rm sat}$ and the star reaches a thermal steady state where viscous heating equals neutrino cooling.
The star then spins down until it reaches the boundary of the instability region, where the amplitude decays.
Different curves on these plots show the results of the evolution for different values of the saturation amplitude, from $10^{-4}$ to $1$. Due to the stronger reheating the curves for larger values of $\alpha_{\rm sat}$ lie at higher temperatures.

On the right panel of Fig.~\ref{fig:alpha-and-T} the evolution of the temperature of the star versus time is shown. Initially the r-mode is absent and the star follows a straight line in the logarithmic $t$-$T$ plot corresponding to mere cooling due to neutrino emission. As can be seen, at large saturation amplitudes $\alpha_{\rm sat} \gtrsim 0.1$ the star temporarily cools below the corresponding steady state curve before the amplitude is large enough so that reheating becomes relevant. Yet, once the r-mode is saturated the dissipative heating slows the cooling of the star significantly at all saturation amplitudes. 
After the star leaves the instability region and the amplitude goes to zero the star cools further by neutrino emission and approaches in the logarithmic plot the initial linear behavior, so that the r-mode effectively merely delays the cooling evolution.
The effect of reheating is even larger for smaller values of $\alpha_{\rm sat}$ because in those cases the spin-down takes longer and the star spends more time in the unstable region, increasing the delay effect.

The dissipation due to bulk viscosity eq. (\ref{eq:bulk-dissipation}) and the neutrino luminosity eq. (\ref{eq:neutrino-emissivity}) were given in the subthermal approximation where non-linear amplitude-dependent terms are neglected. Taking into account the suprathermal non-linear amplitude-dependence \cite{Alford:2010gw} hardly affects the dissipation when there is another non-linear mechanism that saturates the mode due to even stronger dissipation. Although less obvious, we also find that the amplitude dependence of the neutrino luminosity \cite{FloresTulian:2006fq,Jaikumar:2010gf} hardly affects the results since the strong reheating due to the dissipative saturation mechanism strongly delays the cooling and keeps the star at larger temperatures where suprathermal effects are small. This point will be discussed in more detail elsewhere.

To analyze the reheating effect due to the r-modes in detail we compare in Fig.~\ref{fig:T-ratio} the thermal evolution with and without r-mode reheating. Shown is the ratio $T_{\rm r-mode}$ over $T_{\rm no\_r-mode}$, where $T_{\rm no\_r-mode}$ describes the cooling in the absence of r-mode oscillations, versus time in units of the r-mode spin-down time $t_{\rm r-mode}$, which is the time when the star leaves the instability region.
The plot shows that the reheating increases the temperature during the r-mode phase by a constant factor which can take values up to 5 for realistic amplitudes. Whereas for large amplitudes the evolution in the vicinity of the instability region is affected by the simplified model prescription to keep the amplitude constant until the boundary is reached (where it should actually vanish), for sufficiently low amplitudes $\alpha_{\rm sat} < 0.1$  the thermal evolution outside of the instability region takes in the scaled coordinates of Fig.~\ref{fig:T-ratio} a universal form that is independent of the amplitude.
The temperature decays with time scales that are considerably longer than the decay of the r-mode amplitude so that effects of the previous r-mode phase on the thermal evolution are still present long after the r-mode oscillation ceased. This is interesting since, nearly all observed young pulsars spin already too slow that they could be unstable to r-mode oscillations at present, as discussed in more detail in a forthcoming publication \cite{Alford:2012forthcoming}. There we show that r-modes are a consistent mechanism to explain these low frequencies for saturation amplitudes $0.001<\alpha_{\rm sat} < 0.1$ for which r-modes are present over a significant fraction of the age of young stars. Fig.~\ref{fig:T-ratio} shows that even after twice the r-mode spin-down time, the temperature of the star that had an r-mode history is still more than ten percent larger than that of a star without. This presents a unique signature of a previous r-mode phase that is nearly independent of the unknown saturation amplitude.
Although this effect is at present overshadowed by larger observational and model uncertainties, more robust data and a better understanding of these systems could in the future allow to detect a past r-mode phase in temperature measurements of young pulsars.


\begin{theacknowledgments}
MGA thanks the organizers of the QCD@Work 2012 conference for their
hospitality. This research was supported in part by the Offices of
Nuclear Physics and High Energy Physics of the U.S. Department of
Energy under contracts \#DE-FG02-91ER40628, \#DE-FG02-05ER41375. 
\end{theacknowledgments}

\newcommand{\apjl}{Astrophys. J. Lett.\ }
\newcommand{\mnras}{Mon. Not. R. Astron. Soc.\ }
\newcommand{\aap}{Astron. Astrophys.\ }

\bibliographystyle{aipproc}   


\bibliography{cooling_proceeding}

\IfFileExists{\jobname.bbl}{}
 {\typeout{}
  \typeout{******************************************}
  \typeout{** Please run "bibtex \jobname" to optain}
  \typeout{** the bibliography and then re-run LaTeX}
  \typeout{** twice to fix the references!}
  \typeout{******************************************}
  \typeout{}
 }

\end{document}